\documentclass{article}
\usepackage{spconf,amsmath,graphicx,hyperref}
\usepackage{graphicx}
\usepackage{booktabs}
\usepackage{cite}
\usepackage{amsmath,amssymb,amsfonts}
\usepackage{algorithmic}
\usepackage{graphicx}
\usepackage{textcomp}
\usepackage{makecell}
\usepackage{xcolor}

\title{MBCodec: Thorough Disentanglement for High-Fidelity Audio Compression}

\name{Ruonan Zhang\textsuperscript{1},
Xiaoyang Hao\textsuperscript{2}, 
Junjie Cao\textsuperscript{1}, 
Yichen Han\textsuperscript{2},
Yue Liu\textsuperscript{2}, 
Kai Zhang\textsuperscript{1,*}\thanks{No funding was received for conducting this study. The authors have no relevant financial or nonfinancial interests to disclose.}}
\address{\textsuperscript{1}Tsinghua University,
\textsuperscript{2}AMAP Speech}
\begin{document}
%
\maketitle
\begin{abstract}
High-fidelity neural audio codecs in Text-to-speech (TTS) aim to compress speech signals into discrete representations for faithful reconstruction. However, prior approaches faced challenges in effectively disentangling acoustic and semantic information within tokens, leading to a lack of fine-grained details in synthesized speech. In this study, we propose MBCodec, a novel multi-codebook audio codec based on Residual Vector Quantization (RVQ) that learns a hierarchically structured representation. MBCodec leverages self-supervised semantic tokenization and audio subband features from the raw signals to construct a functionally-disentangled latent space. In order to encourage comprehensive learning across various layers of the codec embedding space, we introduce adaptive dropout depths to differentially train codebooks across layers, and employ a multi-channel pseudo-quadrature mirror filter (PQMF) during training. By thoroughly decoupling semantic and acoustic features, our method not only achieves near-lossless speech reconstruction but also enables a remarkable 170x compression of 24 kHz audio, resulting in a low bit rate of just 2.2 kbps. Experimental evaluations confirm its consistent and substantial outperformance of baselines across all evaluations.
\end{abstract}
\section{Introduction}
\label{sec:intro}

Neural audio codecs have consistently attracted significant research attention, universally featuring an encoder-quantizer-decoder framework. High-quality discrete audio codecs are crucial for effectively decomposing speech into distinct subspaces like content, timbre, and prosody\cite{Chen_2022,défossez2024moshispeechtextfoundationmodel}. As the quantizer represents the most critical part of improving codec performance, recent studies have introduced various quantizers, including RVQ\cite{1171604}, GVQ\cite{5432202}, and GRVQ\cite{yang2023hificodecgroupresidualvectorquantization}. However, a key limitation is the lack of effective disentanglement\cite{DBLP:journals/corr/abs-2104-00355} of different speech attributes into distinct representations, which can degrade the overall audio signal quality.

Single codebook approaches such as StableCodec\cite{parker2024scalingtransformerslowbitratehighquality}, XCodec\cite{ye2024codecdoesmatterexploring}, and BiCodec\cite{wang2025sparkttsefficientllmbasedtexttospeech}, present a fundamental trade-off: to achieve a low bitrate, one must increase the code length, which results in an exponential increase in both codebook size and the associated computational burden. In contrast, multiple codebooks architectures, exemplified by SoundStream\cite{zeghidour2021soundstreamendtoendneuralaudio} and EnCodec\cite{défossez2022highfidelityneuralaudio}, address these limitations by sequentially quantizing audio residuals across multiple codebooks. These methods allow for a greater information capacity at a given bitrate. This is because multiple codebooks systems combine vectors multiplicatively ($N^M$
), where $N$ is the codebook size and $M$ is the number of codebooks, while single codebook systems combine them additively ($N\times M$).

However, prior work\cite{gu2024escefficientspeechcoding, wang2023neuralcodeclanguagemodels} on multiple codebook architectures has not adequately addressed a key challenge: the effective disentanglement of semantic and acoustic information from the input signals. This is a critical area of research for enhancing the codec's compression efficiency and reconstruction quality. Secondly, codec training efficiency is constrained by codebook quantity and size, underscoring the need for advancements in training paradigms. Most importantly, the poor interpretability of the individual codebooks in previous codec work led to inefficient model iteration.

This paper proposes MBCodec, a novel codec that transitions RVQ-based codec from conventional residual coding to a functional coding paradigm with explicit training objectives.
\begin{figure*}[h]
  \centering
  \includegraphics[width=0.8\linewidth,keepaspectratio]{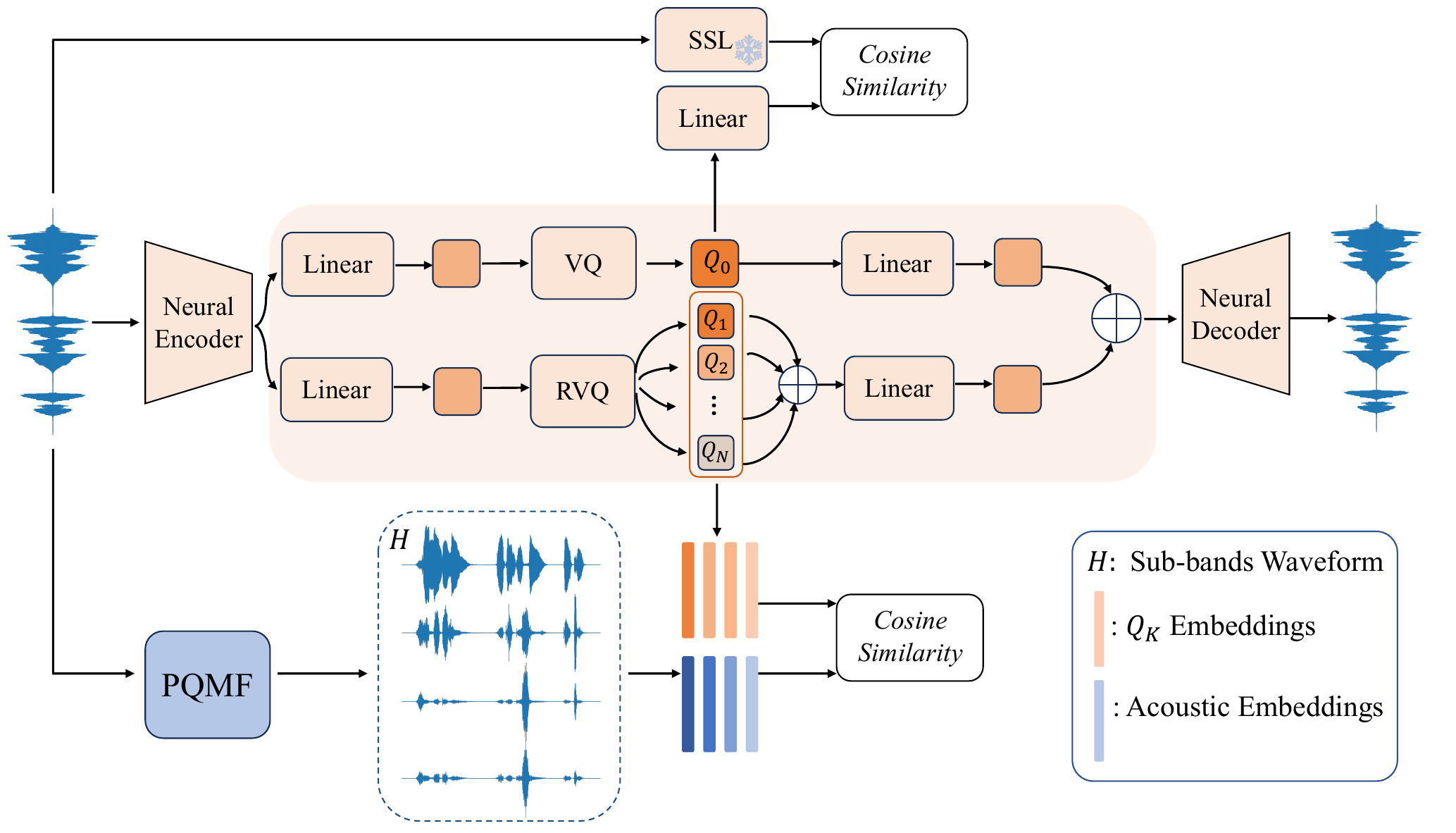}
  \caption{Overall framework of proposed MBCodec.}
  \label{framework}
\end{figure*}
As depicted in Figure\ref{framework}, MBCodec first employs a neural encoder to extract speech embeddings from the input audio. Following this, MBCodec distills semantic information from large-scale Self-Supervised Learning (SSL) models\cite{zhang2024speechtokenizerunifiedspeechtokenizer,DBLP:journals/corr/abs-2110-13900} into a VQ and applies an RVQ in parallel. Additionally, effective supervised acoustic information is vital. We leverage a PQMF filter bank to provide robust, subband-level acoustic supervision for the RVQ codebooks and assign each codebook a distinct physical meaning and function. This design simultaneously boosts model interpretability and overall codec performance. 

Furthermore, MBCodec is trained with an adaptive dropout depths mechanism that dynamically adjusts the number of chosen codebooks, thereby enabling flexible bitrate selection and improving training efficiency. Experimental results demonstrate that MBCodec can reconstruct high fidelity audio at various frame rates with an ultra-high compression ratio that allows it to compress 24KHz audio to an ultra-low bitrate of 2.2kbps. And it also achieves superior performance over baselines in evaluations.

\section{MBCodec}
\label{sec:format}
\begin{table*}[!t]
    \centering
    \caption{Objective evaluation metrics for comparison with baseline codecs.}
    \label{tab:2}
    \begin{tabular*}{\textwidth}{@{\extracolsep{\fill}}lccccccccccc@{}}
        \toprule
        Tokenizer & N & FR & BPS & PESQ $\uparrow$ & SI-SDR $\uparrow$ & STFT $\downarrow$ & Mel $\downarrow$ & MUSHRA$\uparrow$ & WER $\downarrow$ & SIM $\uparrow$ \\
        \midrule
        Ground Truth       & 1  & 24khz     & 384k   & 4.64  & 10.56 & 0.00 & 0.00 &90.7& 3.01 & 0.96   \\
        \midrule
        MBCodec
        
        & 8  & 25Hz & 2.2k   & 2.98 &7.70&0.17 & 3.62 &82.8 &4.25 &0.90\\
         & 8 & 50Hz &  4.4k   & 3.07  &7.72   & 0.12   &3.36 &83.2  &3.38   &0.91 \\
        & 16 & 25Hz & 4.4k   & 3.64  & \textbf{7.98}  & 0.11 & 3.21 &85.3 &3.32 & 0.93\\
       
        & 16 & 50Hz & 8.8k &\textbf{3.83} & 7.94  & \textbf{0.08} & \textbf{2.34} &\textbf{85.9}&\textbf{3.24} & \textbf{0.97}\\
        \midrule
        DAC          
        & 8 & 25Hz & 2.0k   & 3.18  & 7.95 & 0.14 & 5.02 &82.3 & 4.39 &0.83 \\
        
        \midrule
        SpeechTokenizer    & 8  & 50Hz & 4.4k   & 1.26 & 6.66  & 0.58 & 7.02 & 79.0 &5.26&0.82\\
        \midrule
        Encodec & 16 & 50Hz & 8.8k   & 2.78  & 6.48 & 0.36 & 2.96 &85.3 & 4.22 &0.79 \\
        \bottomrule
    \end{tabular*}
\end{table*}
As illustrated in Figure\ref{framework}, MBCodec first employs a neural encoder to compress a 24 kHz audio signal into a sequence of embeddings. These embeddings are then processed by two parallel quantizers: a VQ for semantic information and an RVQ for full-band acoustic information. Finally, a decoder is employed to jointly reconstruct the audio.
\vspace{-2mm}
\subsection{Disentanglement of semantic and acoustic information}

The fidelity of speech reconstruction hinges on the successful recovery of both semantic and acoustic information from the original waveform. This requirement is particularly challenging given that acoustic information is not spectrally uniform, but is heterogeneously distributed across different frequency bands. The low-to-mid frequencies contain the bulk of energy and core intelligibility cues (e.g., fundamental frequency, vocalic formants), while high frequencies encode finer phonetic details like fricatives and sibilants. 

MBCodec achieves this disentanglement through a principled, two-part architecture. First, it enforces robust semantic feature learning by employing direct supervision from a mature HuBERT\cite{DBLP:journals/corr/abs-2106-07447} ASR model. Cocurrently, it captures layered acoustic details by introducing a novel PQMF mechanism. This mechanism decomposes the audio signal into distinct frequency subbands and uses each subband to supervise the training of a corresponding codebook. The PQMF filter bank's design is key to this process. Firstly, the prototype filter, denoted as $h(z)$, is specifically designed as a linear-phase spectral factor of a 2M-band filter $F(z)$. 

\begin{equation}
F(z) = h(z)h(z^{-1})
\label{eq1}
\end{equation}
\vspace{-2mm}

The PQMF filter bank, based on refined cosine modulation and targeted phase selection, is precisely designed to ensure the creation of clean, isolated frequency subbands. We extend Equation~\eqref{eq2}, where the $k_{th}$ frequency subband filter, $h_k[n]$ is derived from a prototype filter $h[n]$, via cosine modulation with the phase term $\phi_k$ = $(-1)^k \frac{\pi}{4}$:

\begin{equation}
h_k[n] = 2 h[n] \cos\left(\frac{\pi}{M}\left(k + \frac{1}{2}\right)n + \phi_k\right)
\label{eq2}
\end{equation}

For each subband, we first apply a Short-Time Fourier Transform (STFT) to obtain its time-frequency representation. This representation is then used to exclusively train a corresponding codebook (e.g., subband signal $k_{th}$ guides codebook $k_{th}$). This targeted training ensures each codebook focuses solely on its assigned frequency range and prevents unnecessary inter-band interference, effectively disentangling the layered acoustic information. Notably, this design obviates the need for introducing additional SSL models to acquire acoustic features. 
\subsection{Adaptive dropout strategy}
\label{adaptive}

To simultaneously boost computational efficiency and encourage robust information disentanglement, MBCodec employs an adaptive dropout strategy. Unlike prior work~\cite{zeghidour2021soundstreamendtoendneuralaudio}, which employs a uniform quantizer dropout strategy by randomly sampling the number of active layers $n_q \in [1, N_q]$, our approach recognizes a mismatch in this design. The nature of RVQ dictates that residual information diminishes with each successive layer. Consequently, later RVQ stages contribute only minor additional information, making this uniform sampling approach suboptimal. 

To address this mismatch, we propose a non-uniform sampling mechanism that better aligns with the hierarchical nature of RVQ. Specifically, we evaluated three candidate distributions: an exponential decay, which strongly prefers leading layers to focus on the most critical information; a half-Gaussian distribution, providing a gentler initial decay rate that balances information from all layers; and a Chi-squared ($\chi^2$), a non-monotonic sampling strategy that avoids over-relying on the leading VQ layers. 
This systematic approach is crucial for understanding the impact of various sampling policies on model training and performance.

Our training strategy unfolds in three sequential stages to ensure both stability and performance. Firstly, a uniform sampling policy is applied across all $N_q$ layers. Once initial convergence is achieved, the process then transitions to the non-uniform sampling policy. Finally, we sample the leading four supervised layers ($n_q \in [1, 4]$) to enhance audio fidelity.
\subsection{Training object}

The entire system is trained using a Generative Adversarial Network (GAN)-based approach. The comprehensive training criteria primarily include adversarial loss $L_{GAN}$, reconstruction loss $L_{recon}$, and codebook losses $L_{vq}$. To achieve a robust disentanglement of semantic and acoustic information, we employ semantic guidance to inform the vector quantization process, as defined by the following equation~\eqref{eq3}:
\begin{equation}
\mathcal{L}_{\text{seman}} =  \cos(\mathbf{Q_{\text{seman}}}, \mathbf{S}),
\label{eq3}
\end{equation}

where $\mathbf{Q}_{\text{seman}}$ and $\mathbf{S}$ denote the quantized output of the VQ and the semantic representation from HuBERT, respectively. The function $\cos(\cdot)$ represents cosine similarity. A key component of our training objective is the acoustic loss, which is supervised by frequency subbands generated by the PQMF analysis, as shown in Equation~\eqref{eq4}:

\begin{equation}
\mathcal{L}_{n_q, \text{acous}} =  \cos(\mathbf{A_1}\mathbf{Q_{\text{acous}}}^{(:,d)}, \mathbf{A_2}\mathbf{H_{\text{i}}}^{(:,d)}),
\label{eq4}
\end{equation}

where $\mathbf{A}_1$ and $\mathbf{A}_2$ denote the projection matrices that map the subbands and codebook representations to a common dimension. The superscript $(:,d)$ denotes the vector for the $d$-th dimension across all time steps. In this context, $\mathbf{Q}_{\text{acous}}$ denotes the acoustic embedding yielded by the ${n_q}$ quantizer, while $\mathbf{H}_i$ represents its corresponding target, the $i_{th}$ frequency subband. The training objective of MBCodec is shown in Equation~\eqref{eq5}:

\begin{equation}
\mathcal{L}_{total}=\mathcal{L}_{GAN} +\mathcal{L}_{recon}+\mathcal{L}_{vq}+\mathcal{L}_{\text{seman}} +\sum_{n_q=1}^{n} \mathcal{L}_{n_q, \text{acous}}
\
\label{eq5}
\end{equation}

where $\mathcal{L}_{\text{seman}}$ and $\mathcal{L}_{n_q, \text{acous}}$ are applied to the RVQ layers to guide the learning of semantic and acoustic properties, respectively. The specific value of $n$ for each training iteration is not predetermined, but is sampled according to the adptive dropout strategy detailed in Subsection~\ref{adaptive}.
\section{experiment}
\label{experiment}

\subsection{Experiment setup}

We benchmark MBCodec against a Descript Audio Codec (DAC)-style baseline \cite{kumar2023highfidelityaudiocompressionimproved}. 
We implement and evaluate two variants of MBCodec with 8 and 16 codebooks respectively, and test each variant at frame rates of both 25 Hz and 50 Hz. Each codebook contains 2048 entries.
Our base models are trained on approximately 95k hours of audio from the Emilia\cite{he2024emiliaextensivemultilingualdiverse} multilingual speech dataset, which is uniformly resampled to 24 kHz. We evaluate the speech reconstruction performance of our models on the test set of AudioSet~\cite{gemmeke2017audio}, following the experimental protocol of DAC.

Our model implementation is detailed as follows. Our model, featuring a 64-dim encoder and a 1536-dim decoder, is trained for two days on a single NVIDIA H20 GPU. During training, we use the snake activation function and a constant learning rate of $1 \times 10^{-4}$. All loss components are equally weighted, and other hyperparameters follow the DAC baseline. For the non-uniform sampling distributions, we use empirically validated parameters, which are detailed in Table~\ref{tab:dist_quality}. These include the base for exponential decay, sigma for the half-Gaussian, and the degrees of freedom for Chi-squared.
\vspace{-4mm}
\subsection{Evaluation on reconstruction signal}

As presented in Table~\ref{tab:2}, our proposed MBCodec is comprehensively compared against the baseline codecs DAC \cite{kumar2023highfidelityaudiocompressionimproved}, Encodec \cite{défossez2022highfidelityneuralaudio} and SpeechTokenizer \cite{zhang2024speechtokenizerunifiedspeechtokenizer}. Notably, our approach achieves its best performance in its 16-codebook, 50 Hz frame rate (FR) configuration, outperforming all competing baseline methods.
Furthermore, our model achieves a high compression ratio of 170x on 24 kHz audio when configured with 8 codebooks at a 25 Hz frame rate. 
\vspace{-2mm}
\begin{table}[!ht]
    \centering 
    \caption{Reconstruction quality under different adaptive dropout distributions.}
    \label{tab:dist_quality}
    \begin{tabular}{lccccc}
        \toprule
        Distribution & PESQ $\uparrow$  & SI-SDR $\uparrow$ & STFT $\downarrow$ & Mel $\downarrow$  \\
      \midrule
        \makecell[l]{Exponential \\ (base=0.6)}
        & 3.02  & 7.61   & 0.34  & 3.93      \\
        \midrule
        \makecell[l]{Half-Gaussian \\ (sigma=5.0)}
        & \textbf{3.28} & \textbf{7.70 }  & \textbf{0.17}  & 3.32   \\
        \midrule
        \makecell[l]{Chi-squared \\ (df=4)}
        & 2.75  & 7.32   & 0.23 & \textbf{3.12}  \\
        \bottomrule
    \end{tabular}
\end{table}

Within our MBCodec variants, models with fewer codebooks (N) consistently yield poorer performance metrics. However, at comparable bitrates, increasing N from 8 to 16 substantially boosts key metrics like PESQ and STFT, demonstrating that a larger codebook size is beneficial for improving audio reconstruction.
Under the same training and experimental settings, we evaluate various adaptive dropout strategies, as shown in Table \ref{tab:dist_quality}. The Half-Gaussian distribution proved to be the most effective, enabling the model to achieve superior reconstruction.
\vspace{-2mm}
\subsection{Visual analysis}

\begin{figure}[htbp]
  \centering
  \includegraphics[width=\columnwidth]{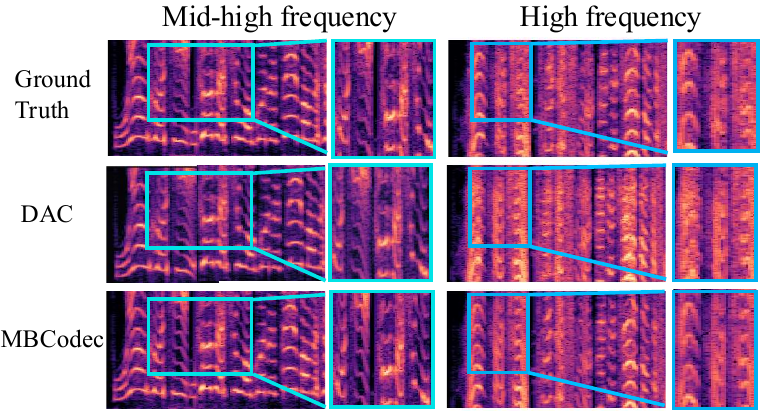}
  \caption{Visualization of the reconstructed spectra of GT, DAC, and MBCodec.}
  \label{FIG2}
\end{figure}

A common bottleneck for neural audio codecs is the challenge of high-frequency reconstruction. As illustrated in Figure \ref{FIG2}, DAC's reconstruction in the mid-high band (2-6 kHz) is noticeably diffuse and lacks spectral definition. This degradation intensifies in the high-frequency region (6-9 kHz), where the output collapses into unstructured noise.
 In sharp contrast, MBCodec maintains high fidelity across the spectrum, faithfully restoring the original signal's spectral structure. This confirms MBCodec's ability to preserve essential harmonics while avoiding the high-frequency artifacts that cause audible distortion. 
\vspace{-2mm}
\vspace{-5mm}

\subsection{Ablation study}

We conduct an ablation study to investigate the contribution of each key component to the final performance. As shown in Table~\ref{tab:3}, removing the PQMF supervision module leads to a significant degradation in PESQ and Mel-spectrogram distance, which measure subjective audio quality and spectral feature similarity. Furthermore, excluding the adaptive dropout function impacts the SI-SDR and STFT metrics, which measure fine-grained frequency changes and high-frequency information. Therefore, both components are indispensable for achieving high-fidelity speech reconstruction.
\vspace{-8mm}
\begin{table}[!ht]
    \centering
    \caption{Results of ablation study.}
    \label{tab:3}
    \begin{tabular}{lccccc}
        \toprule
        Distribution & PESQ $\uparrow$ & SI-SDR $\uparrow$ & STFT $\downarrow$ & Mel $\downarrow$ \\
        \midrule
        MBCodec & \textbf{3.83} & \textbf{7.94} & \textbf{0.08} & \textbf{2.34} \\
        \cmidrule(l){1-5}
        w/o PQMF & 2.34 & 7.69 & 0.19 & 3.45 \\
        \cmidrule(l){1-5}
        \makecell[l]{w/o adaptive \\ dropout} & 3.23 & 7.32 & 0.22 & 3.21 \\
        \bottomrule
    \end{tabular}
\end{table}

\vspace{-5mm}
\section{Conclusion}

We introduce MBCodec, a novel audio codec employing multiple codebooks to explicitly disentangle semantic and acoustic information. This disentangled latent space simultaneously improves compression efficiency for higher-fidelity reconstruction at the same bitrate and enhances model interpretability. To further enhance performance, MBCodec introduces an adaptive dropout mechanism specifically designed to improve reconstruction quality in low bitrate scenarios.
As evidenced by our experimental results, MBCodec not only faithfully preserves the foundational harmonics but also maintains the resolution of high-frequency details essential for perceptual richness. This robust performance ensures nearly lossless reconstruction quality at extremely low bitrates. Additionally, this thorough disentanglement of semantic and acoustic information paves the way for flexible downstream applications like speech synthesis.


\vfill\pagebreak

\bibliographystyle{IEEEbib}
\bibliography{strings}

\end{document}